\documentclass[useAMS,usenatbib]{mn2e}

\usepackage{graphicx, longtable, lscape}

\def\kms{~km~s$^{-1}$}
\def\etal{~et\ al.\ }
\def\h2o{H$_{2}$O}
\def\arcdeg{$^{\circ}$\hspace{-2pt}}
\def\arcmin{$^{\prime}$}
\def\arcsec{$^{\prime\prime}$}

\title[\h2o\ masers in the LMC]{ATCA survey of \h2o\ masers in the Large Magellanic Cloud}
\author[H.~Imai \etal]
{H.~Imai$^{1,2}$\thanks{E-mail: hiroimai@sci.kagoshima-u.ac.jp}, 
Y.~Katayama$^{1}$, S.~P. Ellingsen$^{3}$ and Y.~Hagiwara$^{4,5}$ \\
$^{1}$Department of Physics and Astronomy, Graduate School of Science and Engineering, Kagoshima University, 
1-21-35 Korimoto, \\ Kagoshima 890-0065, Japan \\
$^{2}$International Centre for Radio Astronomy Research, M468, The University of Western Australia, 
35 Stirling Hwy, Crawley, \\ Western Australia, 6009, Australia \\
$^{3}$School of Mathematics and Physics, University of Tasmania, Private Bag 37, Hobart, Tasmania 7001, Australia \\
$^{4}$National Astronomical Observatory of Japan, 2-21-1, Osawa, Mitaka, Tokyo 181-8588, Japan \\
$^{5}$Department of Astronomical Science, The Graduate University for Advanced Studies (Sokendai),\\ 2-21-1, Osawa Mitaka, 
181-8588 Tokyo, Japan
}

\begin{document}

\date{Accepted 2013 February 20. Received 2013 February 19; in original form 2013 February 13}

\pagerange{\pageref{firstpage}--\pageref{lastpage}} \pubyear{2013}

\maketitle

\begin{abstract}
We have analysed archival data taken with the Australia Telescope Compact Array (ATCA) during 2001--2003 and detected nine new interstellar and circumstellar \h2o\ masers in the LMC. This takes the total number of star formation \h2o\ masers in the LMC to 23, spread over 14 different star forming  regions and three evolved stars. Three \h2o\ maser sources (N105a/MC23, N113/MC24, N157a/MC74) have been detected in all the previous observations that targeted these sites, although all show significant variability on timescales of decades. The total number of independent \h2o\ maser sources now known in the LMC means that through very long baseline interferometry astrometric measurements it will be possible to construct a more precise model of the galactic rotation of the LMC and its orbital motion around the Milky Way Galaxy. 
\end{abstract}


\begin{table*}
 \caption{Setup status of the analysed data of the ATCA observations towards the LMC.}
 \label{tab:setup}
\begin{tabular}{llc@{ }cl@{ }c@{ }cl} \hline 
Program & Date & & & & Bandwidth & & \\
code & (yyyy/mm/dd) & $N_{\rm b}$\footnotemark[1] & Array\footnotemark[2] & Scan time\footnotemark[3] & (MHz) & $N_{\rm spect}$\footnotemark[4] & Targets  \\ \hline
C901\ \dotfill \ & 2001/01/07 & 3 & 750C &18--19 $\times$ 3--4 & 8 & 1024 & N105a, N113, N157a, N159, N160a, and 0506$-$612 \\
C973\ \dotfill \ & 2002/11/01--07 & 3 & Wnnn\footnotemark[5] & 1--3$\times$ 2--3 & 16 & 513 & 186 sources in H{\rm II} regions and O-rich AGB stars \\
\hspace*{5mm}\dotfill \ & 2003/05/01--04 & 6 & EW352 & 5--7$\times$ 2--3 & 16 & 513 & 28 sources in H{\rm II} regions and O-AGB 815 \\
\hline
\end{tabular}
\flushleft
\noindent
\footnotemark[1]Number of scans $\times$ duration per scan in min.
\footnotemark[2]Array configuration. 
\footnotemark[3]Number of available baselines.
\footnotemark[4]Number of spectral channels. 
\footnotemark[5]The array configurations are named as combination of antenna positions labelled as W plus three digit numbers such as: W100, W110, W147, W168 and W196. 
\end{table*}

\section{Introduction}

There are more than a thousand \h2o\ maser sources currently known in the Milky Way (MW) galaxy (e.g., \citealt{val01,wal11}). In contrast, the total number of maser sources detected in all species in the Large and Small Magellanic Clouds (LMC, SMC) is around 20 (see \citealt{ell10}, hereafter EBCQF10, and references therein). The maser sources in the LMC and the SMC are important objects for addressing a range of astrophysical questions. 

\h2o\ masers are typically associated with outflow activity from young stellar objects (YSOs) and the final copious stellar mass loss phase of red giant and supergiant stars. All of the \h2o\ masers found in the Magellanic Clouds (MCs) to date are associated with either massive YSOs or red supergiants, both of which trace the present sites of star formation in these galaxies. Taking into account the relatively low metallicity of the MCs compared to the MW, some differences are expected in the dominant physical mechanisms and the timescales that govern the process of star formation and stellar mass loss. 

Very long baseline interferometric (VLBI) observations can resolve each maser source into a cluster (or clusters) of compact maser features, whose three-dimensional motions (line-of-sight velocities and proper motions) can be measured. With current instruments with which astrometric accuracy of 10 micro-arcseconds ($\mu$as) is achievable, such studies are possible for star forming regions in nearby galaxies at distances of up to 1~Mpc (e.g. \citealt{bru05}). Measuring the motions of the maser sources with respect to nearby (in terms of angular separation) quasars, it is possible to determine both a space motion and a trigonometric parallax distance (given sufficient astrometric accuracy) of the maser source. Within the MW on kilo-parsec scales, \h2o\ maser sources are confined to the MW thin disk and have been used to derive fundamental Galactic structure parameters \citep{rei09,hon12}. The LMC and SMC are at distances of approximately 50 and 60~kpc respectively (e.g., \citealt{cio00}), so the amplitude of the trigonometric parallax is approximately 20~$\mu$as. This is too small to enable accurate parallax distance determinations with current instruments and techniques (e.g. \citealt{rei09}), but may be feasible through statistical approaches, or with future space-based VLBI missions. 

With current, ground-based VLBI we are able to determine the space motion of maser sources within the MCs by measuring and correcting for the intrinsic internal motions of maser features in the source. The locations and past orbits of the MCs are important and controversial topics in studies of the star formation history and process of interactions of the galaxies in the MW--MC system (e.g., \citealt{dia12}).  Astrometric observations using data from ground-based optical telescopes and the {\it Hubble Space Telescope} have been used to measure the proper motions of the MCs \citep{pia08, vie10, kal13}. However, there exist non-negligible discrepancies among the results. The optical astrometry results are affected by the inclusion of stars from a variety of populations with different dynamical characteristics located along the same lines of sight.  They are also affected by the galactic rotation model adopted (e.g., \citealt{has12}), which is required to extract the centre-of-mass space motions of the galaxies from the observed proper motions.

In order to analyse the 3-D kinematics of the individual \h2o\ maser sources in the MCs and to derive the dynamical parameters of these galaxies, efforts to increase the number of identified \h2o\ masers in these galaxies are important. Statistical analysis of the 3-D kinematics of the maser sources will enable us to compare the kinematic properties of the MCs with those of the MW. In order to construct dynamical models of the MCs, each galactic model requires the present centre-of-mass space motion, the velocity field of the galactic rotation and the distance to the galaxy as free parameters in the model fitting. A large number of maser sources ($\ga$20) are required to make such an analysis feasible. In this paper, we present \h2o\ masers in the LMC detected with the Australia Telescope Compact Array (ATCA), including nine newly discovered since the work of EBCQF10.

\begin{table*}
 \caption{Interstellar 22-GHz \h2o\ masers in the LMC. Some of the source names in the first column are those registered in the data archive. The source coordinates are cited from the latest result from the present work, EBCQF10, \citet{laz02}  or \citet{oli06}. References for the coordinates are, 1:  \citet{laz02}; 2: \citet{oli06}; 3: EBCQF10; 4: Present work. 1-$\sigma$ noise levels on the images (including side lobes) are given in the present results. References and catalogs for the corresponding objects are, GC: \citet{gru09}; 2MASS: Two-micron All Sky Survey Point Source Catalog; 2MASX: 2MASS Extended Source Catalog.}
 \label{tab:LMC-H2O}
\begin{tabular}{lr@{ }r@{ }rr@{ }r@{ }r rr@{ }r l@{ }rl} \hline 
 & \multicolumn{3}{c}{RA (J2000)} & \multicolumn{3}{c}{Dec. (J2000)} & \multicolumn{2}{c}{Velocity} & 
 \multicolumn{1}{c}{Peak} & & \multicolumn{1}{c}{1-$\sigma$} & \\
 & & & & & & & Range & Peak & \multicolumn{1}{c}{flux} &  & \multicolumn{1}{c}{noise} &  \\
Source name \hspace{14mm} 
& $^{\rm h}$ &  $^{\rm m}$ & $^{\rm s}$ & \arcdeg & \arcmin & \arcsec & \multicolumn{2}{c}{(\kms)} 
& \multicolumn{1}{c}{(Jy)} & Ref. &  \multicolumn{1}{c}{(Jy)} & Counterpart stellar objects \\ \hline
IRAS F04521$-$6928\ \dotfill \ & 4 & 51 & 53.90 & $-$69 & 23 & 27.8 & 205--245 & 233.4 & 2.5 & 4 
& 0.08 & GC J045153.29$-$692328.6 \\
H{\rm II} 1107\ \dotfill \ & 4 & 52 & 09.74 & $-$66 & 55 & 21.8 & 250--275 & 273.5 & 0.9 & 4 
& 0.07 & 2MASS J04520916$-$6655223 \\
H{\rm II} 1186\ \dotfill \ & 4 & 54 & 00.10 & $-$69 &  11 & 54.5 & 245--255 & 245.5 & 5.0 & 4 
& 0.08 & 2MASS J04540020$-$6911557 \\
IRAS~05011$-$6815\ \dotfill \ & 5 & 01 & 01.83 & $-$68 & 10 & 28.2 & 263--266 & 264.4 & 0.08 & 3 
& ... & GC J050101.80$-$681028.5 \\ 
N105a/MC23\ \dotfill \ & 5 & 09 & 52.00 & $-$68 & 53 & 28.6 & 252--260 & 254.6 & 1.4 &  1, 3 
& ... & GC J050952.26$-$685327.3 \\
\hspace*{16mm} \dotfill \ & 5 & 09 & 52.74 & $-$68 & 53 &24.9 & 235--260 & 244.2 & 1.1 &  2, 4 
& 0.09 & GC J050952.26$-$685327.3 \\
\hspace*{16mm} \dotfill \ & 5 & 09 & 52.55 & $-$68 & 53 &28.0 & 246--268 & 266.2 & 0.8 &  3 
& ... & GC J050952.26$-$685327.3\\
N113/MC24\ \dotfill \ & 5 & 13 & 17.17 & $-$69 & 22 & 21.6 & 249--286 & 257.1 & 2.0 & 3 
& ... & GC J051317.69$-$692225.0 \\
\hspace*{16mm} \dotfill \ & 5 & 13 & 17.69 & $-$69 & 22 & 27.0 & 245--250 & 249.1 & 0.3 & 1, 3 
& ... & GC J051317.69$-$692225.0 \\
\hspace*{16mm} \dotfill \ & 5 & 13 & 25.31 & $-$69 & 22 & 42.4 & 215--257 & 237.8 & 52.3 & 2, 4 
& 0.11 & 2MASS J05132499$-$6922442 \\ 
\hspace*{16mm} \dotfill \ & 5 & 13 & 25.15 & $-$69 & 22 & 45.4 & 228--264 & 253.1 & 116 & 1, 2, 3 
& ... & 2MASS J05132499$-$6922442 \\
N113a (in N113) \dotfill \ & 5 & 13 & 22.36 & $-$69 & 22 & 38.6 & 235--265 & 238.1 & 3.1 & 4 
& 2.56 & GC J051321.43-692241.5 \\
IRAS~05202$-$6655\ \dotfill \ & 5 & 20 & 16.90 & $-$66 & 52 & 53.3 & 285--300 & 287.0 & 2.0 & 4 
& 0.06 & 2MASX J05201653$-$6652544 \\
NGC~1984\ \dotfill \ & 5 & 27 & 41.32 & $-$69 & 08 & 05.9 & 255--265 & 258.3 & 3.3 & 4 
& 0.07 & 2MASS J05274010$-$6908044 \\
N157a/MC74\ \dotfill \ & 5 & 38 & 46.53 & $-$69 & 04 & 45.3 & 253--271 & 269.0 & 6.5 & 1, 3 
& ... &GC J053848.42$-$690441.6 \\
\hspace*{16mm} \dotfill \ & 5 & 38 & 49.25 & $-$69 & 04 & 42.1 & 263--268 & 267.3 & 0.5 & 3 
& ... & GC J053848.42$-$690441.6 \\
\hspace*{16mm} \dotfill \ & 5 & 38 & 46.80 & $-$69 & 04 & 45.5 & 235--255 & 249.9 & 0.6 & 2, 4 
& 0.09 & GC J053848.42$-$690441.6 \\
\hspace*{16mm} \dotfill \ & 5 & 38 & 52.67 & $-$69 & 04 & 37.8 & 265--269 & 266.9 & 1.4 & 3 
& ... & 2MASS J05385273$-$6904374 \\
N160a\ \dotfill \ & 5 & 39 & 43.83 & $-$69 & 38 & 33.8 & 236--261 & 251.6 & 1.8 & 1, 2, 3  
& ... & GC J053943.82$-$693834.0 \\
\hspace*{7mm} \dotfill \ & 5 & 39 & 38.97 & $-$69 & 39 & 10.8 & 252--260 & 253.1 & 2.5 & 3 
& ... & GC J053939.02$-$693911.4 \\ 
N159\ \dotfill \ & 5 & 39 & 29.23 & $-$69 & 47 & 18.9 & 242--255 & 248.3 & 1.1 & 1, 3 
& ... & GC J053929.21$-$694719.0 \\
N214b\ \dotfill \ & 5 & 39 & 53.51 & $-$71 & 09 & 51.6 & 225--230 & 228.7 & 7.8 & 4 
& 0.09 &  2MASS J05395357$-$7109533 \\
N180\ \dotfill \ & 5 & 48 & 26.44 &  $-$70 & 08 & 48.8 & 130--140 & 139.8 & 0.6 & 4 
& 0.07 & 2MASS J05482622$-$7008502 \\
\hline
\end{tabular}
\flushleft

\caption{Same as Table \ref{tab:LMC-H2O} but for a circumstellar 22-GHz \h2o\ maser in the LMC, which was newly detected in the present work.}
\label{tab:LMC-H2O-AGB}
\begin{tabular}{l r@{ }r@{ }rr@{ }r@{ }r rr@{ }rrl} \hline 
 & \multicolumn{3}{c}{RA (J2000)} & \multicolumn{3}{c}{Dec. (J2000)} & \multicolumn{2}{c}{Velocity} & 
 \multicolumn{1}{c}{Peak} & \multicolumn{1}{c}{1-$\sigma$} \\
 & & & & & & & Range & Peak & \multicolumn{1}{c}{flux} &  \multicolumn{1}{c}{noise} & \\
Source name \hspace{14mm} 
& $^{\rm h}$ &  $^{\rm m}$ & $^{\rm s}$ & \arcdeg & \arcmin & \arcsec & \multicolumn{2}{c}{(\kms)} 
& \multicolumn{1}{c}{(Jy)} & \multicolumn{1}{c}{(Jy)} & Counterpart stellar object \\ \hline
O-AGB 815 & 5 & 35 & 14.57 & $-$67 & 43 & 54.5 & 230--236 & 235.5 & 0.18 
& 0.08 & HV~1001, 2MASS J05351409$-$6743558  \\
\hline
\end{tabular}

\noindent
{\it Note}. The coordinates are cited from the Harvard Variable (HV) Star Catalogue. 
\end{table*}

\begin{figure*}
\includegraphics[width=173mm]{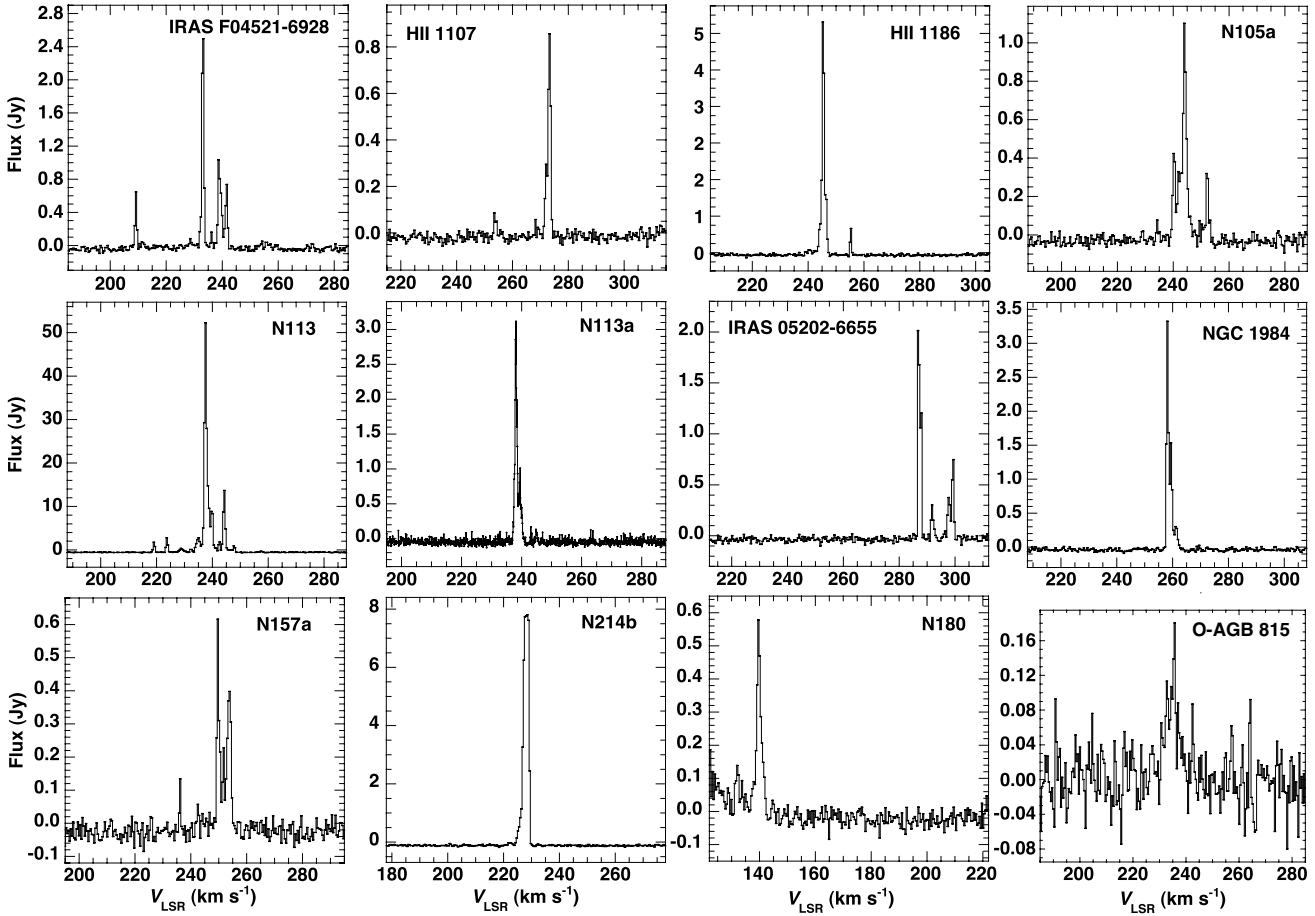}
\caption{\h2o\ maser spectra taken with the ATCA in the 2003 May session, except for N113a taken in the 2002 November session.}
\label{fig:spectra}
\end{figure*}

\section[]{Analysis of ATCA archival data}
\label{sec:analysis}

We analysed archival observations of \h2o\ maser sources in the MCs taken in ATCA observations made on 2001 January 7 (program C901), and in the period of 2002 February--2003 May (C973)\footnote
{The C973 observations also targeted sources in the SMC, which we also reduced. However, the SMC data have been independently analysed and published by \citet{bre13} and in this paper we focus only on the sources in the LMC.}. 
Table \ref{tab:setup} summarises the observing strategy and correlator setup used for the observations. In each of the observing sessions all six 22-m antennas were used in the array, but for some observations only three baselines were useable. Data reduction was undertaken using the {\it MIRIAD} package, following the standard procedures for ATCA data. The visibility amplitude and phases were calibrated by referencing them to observations of flux density and phase calibrator sources. Phase calibration solutions were obtained from the scans on the phase calibrators, J045005.4$-$810102 and J050644.0$-$610941, one of which was observed every 10--20 min. Although there was a difference in the channel resolution for the correlator outputs in the different sessions, all data were smoothed to the achieve the same velocity resolution of $\sim$0.5\kms . We then constructed synthesis images for each source to search for \h2o\ masers and to check the image fidelity. 

For sources where a maser detection was confirmed and there was sufficient coverage of the uv-plane (mainly in the C973 2003-June session), we attempted to obtain the spectrum and the coordinates of the maser emission from the image cube. Even for the sources with sufficient uv-coverage, there  existed significant side lobes, making the astrometry difficult. The angular resolution of the ATCA observations was typically $\sim$0\arcsec.5 (corresponding to a linear resolution of $\sim$0.1~pc at the distance of the LMC), within which most velocity components of a single maser source will be contained. Because we are mainly interested in the number of independent maser sources rather than the internal structures of the individual maser sources in this paper, we have only determined the coordinates of the brightest velocity components of the maser sources. The uncertainty in the measured maser positions is 1--2\arcsec\ in the cases where we could not uniquely determine the brightest point of the maser source due to the high side lobe levels. Columns 2 and 3 of Table \ref{tab:LMC-H2O} give the coordinates of the detected maser sources where we were able to determine them. 

\section{Results}
A total of twelve 22-GHz \h2o\ masers were detected towards the LMC in the C901 and C973 ATCA observations, eight and one of which are newly discovered interstellar and circumstellar sources, respectively. Table \ref{tab:LMC-H2O} lists the sixteen interstellar \h2o\ masers in the LMC known to date (see also the latest review of maser source surveys towards the MCs in \citealt{loo12}). Table \ref{tab:LMC-H2O-AGB} gives the parameters of the circumstellar maser newly detected. Figure \ref{fig:spectra} shows the ATCA cross-power spectra of the \h2o\ masers in the LMC. 

The stellar source O-AGB 815 (2MASS J05351409$-$6743558) corresponds to an M4-type variable star (HV 1001) with a $K-$magnitude of 8.14. Its \h2o\ maser is apparently weak ($\sim$0.2~Jy) and the maser luminosity is comparable to or lower than that seen in supergiants in the MW (e.g., VY CMa with an \h2o\ maser flux density of up to 1000~Jy at a distance of 1.1~kpc, \citealt{cho08}). 

The eight newly detected interstellar \h2o\ masers (those where Column 7 of Table \ref{tab:LMC-H2O} contains only the reference 4) are IRAS\,F04521$-$6928, H{\rm II} 1107, H{\rm II} 1186, N113a (in N113/MC24), IRAS~05202$-$6655, NGC~1984/OH~279.6$-$32.6, N214b and N180. In some regions, \h2o\ maser sources were spread over an area of order 1\arcmin. Taking into account the position uncertainty, we regarded maser emission located within a radius of 1\arcsec--2\arcsec\ as a single source and they are shown in a single line in Table \ref{tab:LMC-H2O}.  There are fifteen independent \h2o\ maser sources that are separated from each other by more than 1\arcmin\ (15~pc at the distance of the LMC). 

Some bright \h2o\ maser sources were first detected more than 25 years ago (e.g., \citealt{sca81}). However, even for regions where there has been persistent maser emission over a long period, the sources have exhibited large flux density variability and sometimes drop below the detection threshold. For example, in the brightest maser source N113/MC24, the peak velocity of the brightest maser component has dramatically changed and the strongest spectral feature observed in this paper (observation in 2003 May) was not detected in the observations of EBCQF10 (2008 August). As a result, apparent position shifts ($>$1\arcsec\ or $>$0.25~pc) may occur due to the maser sources with such flux variability in the different locations (a Christmas tree effect). Clustering of \h2o\ maser sources on parsec scales is observed in some Galactic star formation regions (e.g., W51 North and Main; W3(OH) and W3~IRS5). Therefore, the observed position shifts may be due to variation in the relative flux density between maser sources within a cluster, although we cannot rule out the possibility that these shifts occur just due to the relative positional errors of the observations made at different epochs. N113/MC24 and N113a present one example of a cluster of \h2o\ maser sources (\citealt{laz02,oli06}; EBCQF10). 

We looked for corresponding objects within 10\arcsec\ of the maser sources and these are listed in Column 9 of Table \ref{tab:LMC-H2O}. Only the nearest sources from the maser position are listed. The corresponding 2MASS sources are located within 3\arcsec\ of the masers. Note that IRAS~05202$-$6655 seems to be associated with an extended infrared source (2MASX J05201653$-$6652544) and its population is unclear. Proximity to sources listed by \citet{gru09} may be a good indication of association with a YSO. 

\section{Discussion}

Here we estimate possible detections of \h2o\ masers in the LMC using an \h2o\ maser luminosity function (\h2o-LF). \citet{gre90} derived the \h2o-LF
for the MW, which will be improved by further unbiased \h2o\ maser surveys in the MW (e.g., \citealt{wal11}). Similar to the approach used by \citet{dar11} to estimate the expected number of \h2o\ masers in M31, we need to rescale the \h2o-LF using the relative star formation rates of the MW and the LMC ($\sim$4 and $\sim0.4\: M_{\sun}{\rm yr}^{-1}$, respectively, \citealt{die06,ski12}). 

The shortest on-source integration time for the present data was $\sim$3~min. With a velocity resolution of 0.5\kms, the corresponding detection limit for the peak of any \h2o maser emission is estimated to be 1~Jy.  Adopting a distance to the LMC of 51~kpc (corresponding to a distance modulus of 18.55, e.g., \citealt{cio00,has12}) our detection limit corresponds to an isotropic \h2o maser luminosity of $2.4\times 10^{-5}\:L_{\sun}$. We calculate the expected number of \h2o\ masers with the luminosities higher than this value in the LMC to be $\sim$15. This is in good agreement with the actual number of the identified sources over 1~Jy (16 sources in Table~\ref{tab:LMC-H2O}). This supports the argument that the number of \h2o\ maser detections in the LMC is consistent with MW expectations when the difference in the star formation rates between the LMC and the MW are taken into account (\citealt{gre08}; EBCQF10, and references therein). Although the same approach when applied to M31 results in an underestimation of the expected number of detections (prediction of $\sim$3 compared to the 5 detected, \citealt{dar11}), the difference is less than a factor of 2 and within the estimated uncertainty of this approach.

EBCQF10 used the LMC YSO catalogue of \citet{gru09} to investigate the infrared properties and spectral energy distribution (SED) of star formation regions with an associated \h2o\ maser and compared them to the entire sample.  They showed that \h2o-associated YSOs have higher central mass, larger envelope radius, higher ambient density and higher total luminosity than the general YSO population.  EBCQF10 classified YSO within 2\arcsec of the \h2o\ maser position as being associated.  Applying the same criterion to the 8 new LMC \h2o masers listed in Table~\ref{tab:LMC-H2O}, we find three (H{\rm II} 1186, N214b \& N180) have a definite YSO each within 2\arcsec\ (table 9 of \citealt{gru09}).  In contrast, for the 16 LMC interstellar masers investigated by EBCQF10, 11 were within 2\arcsec\ of a definite YSO identified by \citet{gru09}.  The cause of this discrepancy is not clear, it may indicate that the newly detected LMC \h2o\ maser sources are primarily associated with more evolved star formation regions than the sample of EBCQF10, which targeted all the CH$_3$OH, OH and \h2o\ masers known at that time. The three associations from the newly detected masers have similar SED properties to the maser-associated sources of EBCQF10.  However, none of them are sources identified by EBCQF10 as a high-probability maser candidate, so future searches towards these sources may provide additional \h2o masers in the LMC. 

\begin{figure}
\centering
\includegraphics[width=65mm]{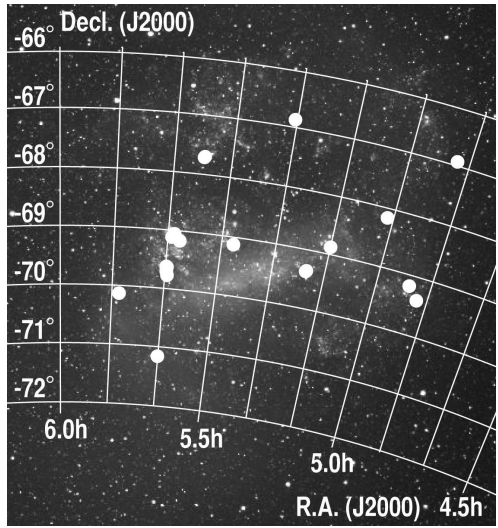}
\caption{Distribution of all interstellar \h2o\ masers identified to date in the LMC on a gnomonic projection map. The background grey image is the optical image of the LMC taken from {\it SkyView}: http://skyview.gsfc.nasa.gov.}
\label{fig:distribution}
\end{figure}

Including the three circumstellar \h2o\ masers associated with red supergiants (IRAS~04553$-$6825, 05280$-$6910, \citealt{loo01}; HV~1001) as members of the young stellar population, there are 19 independent maser sources in the LMC.  Even limited to 16 interstellar masers brighter than 0.5~Jy, through measurement of the proper motions of these masers (better than 100~$\mu$as~yr$^{-1}$ in a time baseline of one year) it is possible to obtain 48 observables (the three-dimensional velocity vector components for each source).  A basic dynamical model of the LMC requires nine free parameters in the fitting: a source distance, a three-dimensional centre-of-mass space motion vector, the galactic rotation axis inclination and position angle and a third (or higher order) polynomial galactic rotation curve. Fig.~\ref{fig:distribution} shows that the \h2o\ masers are distributed throughout the LMC, mitigating the likelihood of bias in the model fitting which may occur if only a small number of independent locations were available. If the distance to the LMC and the axis of the galactic rotation are fixed in the model fitting (using previous accurate determinations), it will be possible to extract the velocity field of the LMC and more precisely determine the centre-of-mass space motion of the LMC from maser proper motion observations.

\section{Conclusion}

With the current sample of \h2o\ masers in LMC (19 in total) and the existence of position-reference quasars within a few degrees of arc of these masers (Imai \etal\ in preparation), VLBI astrometric observations to estimate the centre-of-mass space motion vector for the LMC are feasible. They may  provide an opportunity for independently checking the results of optical astrometry measurements. Furthermore, measurement of the trigonometric parallax of the LMC ($\pi\approx20\:\mu$as ) may be possible. With a position measurement accuracy achieved by the state-of-the-art VLBI astrometry mentioned previously, an individual measurement of the parallax will yield only a marginal detection. However, with independent measurements from $\sim$10 regions, each with potentially a few detections of the maser spot parallax, it will be possible to statistically improve the reliability and precision of the parallax detection. In the same observations it will also be possible to investigate the spatio-kinematical structure for each \h2o\ maser source and to study the physical properties of the star formation regions and the final stellar mass loss in the MCs.

\end{document}